\documentclass{aa}  
\usepackage{graphicx,subfigure,enumerate,pstricks,latexsym,txfonts,natbib,ulem,cancel}
\usepackage[english]{babel}

\newcommand{\beq}{\begin{equation}}
\newcommand{\eeq}{\end{equation}}
\newcommand{\bea}{\begin{eqnarray}}
\newcommand{\eea}{\end{eqnarray}}

\def\tt{\theta}

\def\mir{\mathrm{r}}
\def\mit{\mathrm{\theta}}
\def\mip{\mathrm{\phi}}

\begin{document}

\title{Models of quasi-periodic oscillations related to mass and spin of the GRO~J1655-40 black hole}

\titlerunning{Parameters of the GRO J1655-40 black hole}

\author{Zden\v{e}k Stuchl{\'i}k \inst{1}
          \and
                                Martin Kolo\v{s} \inst{1}
        }
\authorrunning{Z.~Stuchl{\'i}k and M.~Kolo\v{s}}

\institute{
Institute of Physics and Research Centre for Theoretical Physics and Astrophysics, Faculty of Philosophy and Science, 
Silesian University in Opava,  Bezru{\v c}ovo n{\'a}m.13, 
CZ-74601 Opava, Czech Republic
          }

\date{\today}

\abstract
   {}
        {
        Frequencies of the three quasi-periodic oscillation (QPO) modes observed simultaneously in the accreting black hole GRO J1655-40 are compared with the predictions of models. Models in which all  three QPO signals are produced at the same radius are considered: these include different versions of relativistic precession, epicyclic resonance, tidal disruption, and warped disc models. Models that were originally proposed to interpret only the twin high-frequency QPOs are generalized here to interpret also the low-frequency QPO in terms of relativistic nodal precession. Inferred values of the black hole mass and spin from each QPO model are compared with the mass measured from optical observations and the spin inferred from X-ray spectroscopy techniques. We find that along with the relativistic precession model predicting $M=(5.3\pm0.1)~M_{\odot}, a=0.286\pm0.004$,  the so-called total precession model ($M=(5.5\pm0.1)~M_{\odot}, a=0.276\pm0.003$), and the resonance epicyclic model with beat frequency ($M=(5.1\pm0.1)~M_{\odot}, a=0.274\pm0.003$) also satisfy the optical mass test. We compare our results with those inferred from X-ray spectral measurements.
        }
         {}
   {}
   {}


\maketitle

\section{Introduction}\label{intro}

The relativistic precession (RP) ``hot spot'' model \citep{Ste-Vie:1999:PHYSRL:} of twin high-frequency (HF) quasi-periodic oscillations (QPOs) combined with the relativistic nodal model of the low-frequency (LF) QPOs \citep{Ste-Vie:1998:ApJ:} can be  applied well to  the stable twin HF QPOs with $3:2$ frequency ratio observed in the microquasar GRO~J1655-40 simultaneously with the related LF QPO \citep{Mot-etal:2014a:MNRAS:}. Different models of the twin HF QPOs were related to data observed in some other microquasars, giving restrictions on the black hole mass and spin of XTE~J1550-564 \citep{Mot-etal:2014b:MNRAS:}, or GRS 1915+105 \citep{Tor-etal:2005:ASTRA:}. \footnote{For the microquasars GRS 1915+105, GRO~J1655-40, and XTE~J1550-564, the observed twin HF QPOs with $3:2$ frequency ratio cannot be explained by a fixed oscillation model based on frequencies of geodesic quasi-circular motion if we assume central Kerr black holes \citep{Tor-etal:2011:ASTRA:}, while a unique (epicyclic resonance) model exists \citep{Kot-etal:2014:ASTRA:}, if the central objects are Kerr naked singularities that have the special characteristic of corotating circular geodesics \citep{Stu:1980:BAC:,Stu-Sche:2010:CLAQG:}.} The models of twin HF QPOs can also give interesting restrictions  on parameters of neutron stars \citep{Mil-Lam-Psa:1998:ApJ:,Zhang-etal:2006:MONNR:,Bel-Men-Hom:2007:MONNRS:,Muk:2009:ApJ:,Tor-etal:2010:ApJ:,Lin-etal:2011:ApJ:,Tor-etal:2012:ApJ:,Mon-Zan:2012:MNRAS:,Papp:2012:MONNR:,Bosh-etal:2014:GraCos:,Ste:2014:MNRAS:,Stu-etal:2015:ACTA:}. 

Here we test, whether the models matching the twin HF QPOs with frequency ratio $3:2$ in the microquasars XTE J1550-56 and GRS 1915+105 could match the three QPO set observed in the microquasar GRO J1655-40, while we generalize these models to include the relativistic nodal precession. To match the observational data in GRO~J1655-40, we apply a variety of twin HF QPO models based on the frequencies of the geodesic epicyclic motion of matter in accretion discs orbiting Kerr black holes, i.e. the orbital (azimuthal) frequency of the circular motion, or the radial and vertical epicyclic frequencies. The nodal oscillation model based on the Lense-Thirring frequency of the geodesic motion is applied to the LF QPO simultaneously observed with the twin HF QPOs. We thus consider only oscillation models of the twin HF QPOs and the LF QPOs where purely gravity (geometry) of the Kerr black hole is essential.  

We restrict our attention to the models that assume occurrence of the twin HF QPOs (and the simultaneously observed LF QPO) at a common radius. We thus exclude the discoseismic models assuming that the oscillatory modes giving the twin HF QPOs arise at different radii of the accretion disc as they  do not interact and evolve independently \citep{Kat-Fuk:1980:PASJ:,Zan-etal:2005:MNRAS:}. We  study the RP model along with its variants and the epicyclic resonance (ER) model and its variants. We extend this selection for the tidal disruption model where the twin HF QPOs are created by inhomogeneities deformed to a ring by tidal forces of the black hole \citep{Cad-Cal-Kos:2008:ASTRA:,Kos-etal:2009:ASTRA:}, and by the model of warped thin disc oscillations \citep{Kat:2004:PASJ:,Kat:2008:PASJ:}. Frequencies of the twin oscillatory modes used in the twin HF QPO models are given in Tab. \ref{tab1}. 

Contrary to the Monte Carlo technique applied in \cite{Mot-etal:2014a:MNRAS:}, here we use the frequency ratio technique inspired by the resonance conditions relating the dimensionless black hole spin $a$ to the common dimensionless radius $x$ in the resonance models of the twin HF QPOs \citep{Stu-Kot-Tor:2012:ACTA:,Stu-Kot-Tor:2013:ASTRA:}. 

We demonstrate that both a twin HF QPOs model and the nodal precession model imply a mass-spin relation, and the combination of these two relations gives limits on mass and spin of the black hole. The limits on the mass of the GRO~J1655-40 black hole implied by the models are tested by the mass limits obtained from the optical measurements \citep{Bee-Pod:2002:MNRAS:}, while the limits on the spin can be tested by limits from X-ray spectral measurements which might be affected by substantially higher systematics. 

\begin{figure}
\centering 
\includegraphics[width=\hsize]{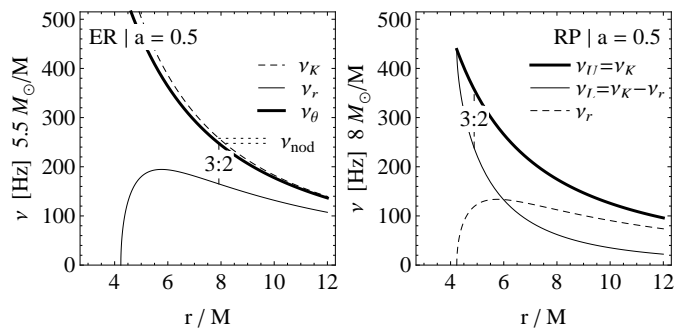}
\caption{
Radial profiles of the orbital, radial, and latitudinal harmonic frequencies $\nu_{\rm \mip}(r)$, $\nu_\mir(r),$ and $\nu_\mit(r)$ related to the static distant observers for particle oscillations relevant for the ER model (left) and their combinations related to the RP model (right).
\label{RFprofiles}
}
\end{figure}

\section{Observational data of GRO~J1655-40}

Mass of the GRO~J1655-40 black hole is estimated by dynamical studies based on spectro-photometric optical techniques \citep{Bee-Pod:2002:MNRAS:} that are not related to the timing studies based on the X-ray measurements, and the range of allowed values of the mass parameter reads 
\beq
     M_{\rm opt} = (5.4\pm0.3)~M_{\odot}. \label{Mopt}
\eeq

The Rossi XTE observatory brings many timing measurements of the X-rays emitted by the GRO~J1655-40 source, which  are summarized in \cite{Mot-etal:2014a:MNRAS:}. The LF QPOs were observed between $0.1Hz$ and $30Hz$ \citep{Cas-Bel-Ste:2005:ApJ:,Mot-etal:2012:MNRAS:}. However, the most important for our study is the simultaneous observation of twin HF QPOs at frequencies $\sim~300$~Hz and $\sim~450$~Hz, and the LF QPO at frequency $\sim~17$~Hz that was reported in \cite{Str:2001:ApJ:}. 

We shall consider here the group of twin HF QPOs and LF QPO presented as Sample B1 in Tab. 2. of \cite{Mot-etal:2014a:MNRAS:}. The lower and upper frequency of the twin HF QPOs and the simultaneously observed LF QPO frequency at the sample read (in Hertz)
\beq
 f_{\rm L} = 298\pm4, \quad f_{\rm U} = 441\pm2, \quad f_{\rm low} = 17.3\pm0.1. \label{ff}
\eeq

We use this set of the peak frequencies of QPOs, taking into account the measurement errors of the peak frequencies (centroid frequencies dominated by statistics of the measurements) to obtain estimates of the GRO~J1655-40 black hole mass $M$ and dimensionless spin $a$. In the following, we use the frequency ratio method developed in \cite{Stu-Kot-Tor:2013:ASTRA:}.

\section{Oscillation models with frequencies governed by geodesic quasi-circular motion}

In the Kerr spacetimes, circular geodesics can exist only in the equatorial plane \citep{Bar-Pre-Teu:1972:ApJ:,Stu:1980:BAC:}. The orbital frequency $\nu_{\rm \mip}$ of the circular geodesic motion, the vertical epicyclic frequency $\nu_{\theta}$, and the radial epicyclic frequency $\nu_{\rm r}$ of the near-circular epicyclic motion are given and discussed in \cite{Ali-Gal:1981:GRG:,Ste-Vie:1998:ApJ:,Tor-Stu:2005:ASTRA:} and {Stu-Sche:2012:CLAQG:}. Radial extension of the quasi-circular geodesic motion has been discussed in \cite{Stu-Kot-Tor:2011:ASTRA:}. 

The hot spot models assume radiating spots in thin accretion discs following nearly circular geodesic trajectories. In the standard RP model \citep{Ste-Vie-Mor:1999:ApJ:}, the upper of the twin frequencies is identified with the orbital (azimuthal) frequency, $\nu_{\rm U}=\nu_{\rm \mip}$, while the lower one is identified with the periastron precession frequency, $\nu_{\rm L}=\nu_{\rm \mip} - \nu_\mir$. The LF QPOs are related to the nodal (Lense-Thirring) precession with frequency $\nu_{\rm nod}=\nu_{\rm \mip}-\nu_\mit$. The radial profile of the frequencies $\nu_U$ and $\nu_L$ of the RP model is presented in Fig. \ref{RFprofiles}. From the variants of the RP model \citep{Stu-Kot-Tor:2013:ASTRA:}, we select the RP1 model introduced in \cite{Bur:2005:RAG:}, where $\nu_{\rm U}=\nu_{\rm \mit}$ and $\nu_{\rm L}=\nu_{\rm \mip} - \nu_\mir$, and the ``total precession model'' RP2 introduced in \cite{Stu-Kot-Tor:2013:ASTRA:}, where $\nu_{\rm U}=\nu_{\rm \mip}$ and $\nu_{\rm L}=\nu_{\rm \mit} - \nu_\mir$ (see Tab. \ref{tab1}). Both the RP1 and RP2 models predict frequencies $\nu_{U}$ and $\nu_{L}$ close to those of the RP model. The combination of the RP model of twin HF QPOs and the nodal model of LF QPO is a fundamental feature of the hot spot kinematic QPO model introduced in \cite{Ste-Vie:1999:PHYSRL:,Ste-Vie-Mor:1999:ApJ:}. Here we apply the assumption of relevance of the nodal frequency model for the LF QPO to both the RP1 and RP2 models of twin HF QPOs -- the frequencies $\nu_{\phi}$ and $\nu_{\theta}$ entering the nodal frequency are involved in both RP1 and RP2 models. 

The tidal disruption (TD) model, where $\nu_{\rm U}=\nu_{\rm \mip} + \nu_\mir$ and $\nu_{\rm L}=\nu_{\rm \mip}$, could resemble to some degree the hot spot models as numerical simulations of disruption of inhomogeneities (e.g. asteroids) by the black hole tidal forces demonstrate existence of an orbiting radiating core in the created ring-like structure \citep{Cad-Cal-Kos:2008:ASTRA:,Kos-etal:2009:ASTRA:}. In order to also apply  the nodal frequency for the LF QPOs, we have to introduce the assumption of vertical oscillatory motion of the distorted inhomogeneity, as the frequency $\nu_{\theta}$ is not included in the TD model. 

The epicyclic resonance (ER) models \citep{Abr-Klu:2001:ASTRA:} consider a resonance of axisymmetric oscillation modes of accretion discs that can be geometrically thin, with geodetical radial profile of angular velocity \citep{Nov-Tho:1973:BlaHol:,Pag-Tho:1974:ApJ:}, or toroidal and geometrically thick, having an angular velocity radial profile governed by gravity and pressure gradients \citep{Koz-etal:1977:ASTRA:,Abr-etal:1978:ASTRA:,Stu-etal:2009:CLAQG:}. Frequencies of the disc oscillations are related to the orbital and epicyclic frequencies of the circular geodesic motion for both geometrically thin discs \citep{Kat-Fuk-Min:1998:BHAccDis:,Kat:2004:PASJ:} and slender toroidal discs \citep{Rez-etal:2003:MNRAS:,Mon-Zan:2012:MNRAS:}. The radial profile of the frequencies $\nu_U$ and $\nu_L$ of the ER model is presented in Fig. \ref{RFprofiles}. In the ER model that has axisymmetric oscillatory modes with frequencies $\nu_{\theta}$ and $\nu_{r}$, the oscillating torus (or circle) is assumed to be radiating uniformly. A sufficiently large inhomogeneity on the radiating torus, which orbits with the frequency $\nu_{\phi}$, enables the introduction of the nodal frequency related to this inhomogeneity. 

The parametric resonance of the radial and vertical epicyclic oscillatory modes is governed by the Mathieu equation predicting the strongest resonant phenomena for the frequency ratio $3:2$ \citep{Lan-Lif:1969:Mech:,Nay-Moo:1979:NonOscilations:}. The forced non-linear resonance admits the presence of combinational (beat) frequencies in the resonant solutions \citep{Nay-Moo:1979:NonOscilations:}. For example, the beat frequency $\nu_{-}=\nu_\mit-\nu_\mir$ implies the observed frequency ratio $\nu_\mit:\nu_{-}=3:2$ at the radius where the frequency ratio $\nu_\mit:\nu_\mir=3:1$ \citep{Stu-Kot-Tor:2013:ASTRA:}. We also define  the beat frequency $\nu_{+}=\nu_\mit+\nu_\mir$ and assume combinations of the epicyclic frequencies with the beat frequencies to give the variants of the ER model. Five additional variants are under consideration and  are summarized in Tab. \ref{tab1}. As in the ER model, we assume the existence of the additional nodal frequency mode due to a torus inhomogeneity for all  five variants. 

The warped disc (WD) oscillation model of twin HF QPOs assumes non-axisymmetric oscillatory modes of a thin disc \citep{Kat:2004:PASJ:,Kat:2008:PASJ:}. For the purposes of the present study, we include again the nodal precession model of the LF QPOs into the framework of the WD model. However, for the WD model with frequencies presented in  Tab. \ref{tab1}., we have to introduce the vertical oscillatory frequency $\nu_{\theta}$ by assumption of vertical axisymmetric oscillations of the thin disc. 

The frequency resonance conditions of the parametric and forced resonances are identical, but the resonant frequency width, resonance strength, and time evolution differ \citep{Nay-Moo:1979:NonOscilations:}. We concentrate  on the resonance frequency conditions only. The present quality of the HF QPO measurements is not sufficient to test the detailed predictions of the parametric or forced resonances. The parametric resonance admits scatter of the resonant frequencies -- the resonance can occur while the oscillating modes in resonance have a frequency ratio  that differs slightly from the exact rational ratio; the width of the resonance scatter strongly decreases with increasing order of the resonance \citep{Lan-Lif:1969:Mech:}. For the forced resonances, the scatter of the frequency ratio from the rational ratio is governed by non-linear effects \citep{Nay-Moo:1979:NonOscilations:}. \footnote{We expect that the LOFT observatory \citep{Fer-etal:2012:ExpAstr:} enables precision of the frequency measurement that is high enough to follow the details of the resonant phenomena.} 

The~resonance condition is given in terms of the~rational frequency ratio parameter $p=\left(\frac{m}{n}\right)^2$ \citep{Stu-Kot-Tor:2013:ASTRA:}. We can use the generalized condition allowing for the resonance scatter, assuming a non-rational ratio of the observed lower and upper frequencies of the twin HF QPOs that is in vicinity of the $3:2$ ratio as given by the frequency measurement errors. Then the~frequency ratio parameter simply reads $p = \left(\frac{\nu_{\rm L}}{\nu_{\rm U}}\right)^2$. 

For further considerations it is useful to introduce the dimensionless radius by the relation $x=r/r_{g}$, where the gravitational radius $r_{g}=GM/c^2$. The resonance (frequency ratio) relations determining the dimensionless radius $x^{\nu_{\rm U}(\mip,r,\theta)/\nu_{\rm L}(\mip,r,\theta)}(a,p)$ where the twin oscillations with the upper (lower) frequency $\nu_{\rm U}(\mip,r,\theta)$ ($\nu_{\rm L}(\mip,r,\theta)$) determined by a concrete twin HF QPOs model occur are presented in \cite{Stu-Kot-Tor:2013:ASTRA:}. 

\begin{figure}
\includegraphics[width=0.49\hsize]{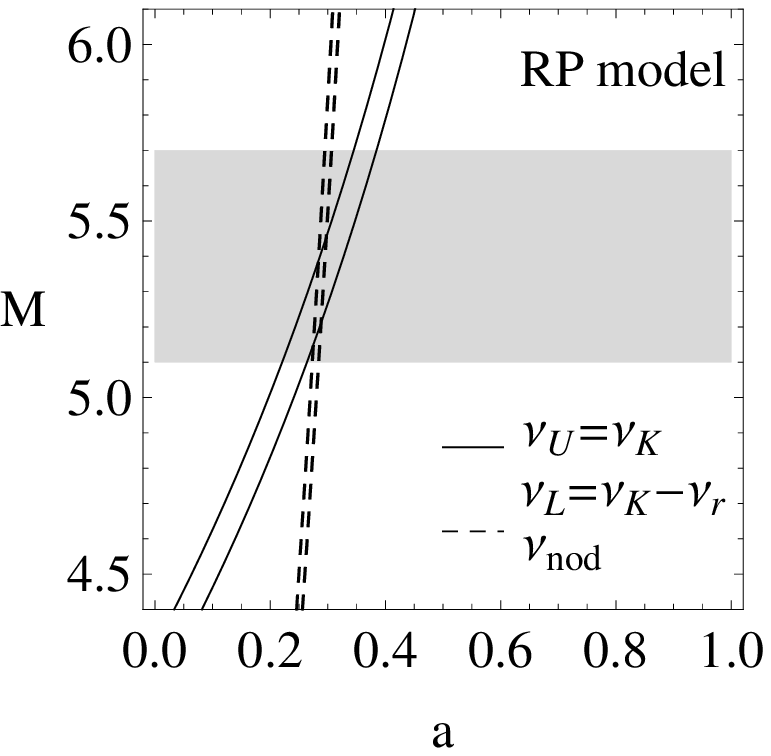}
\includegraphics[width=0.49\hsize]{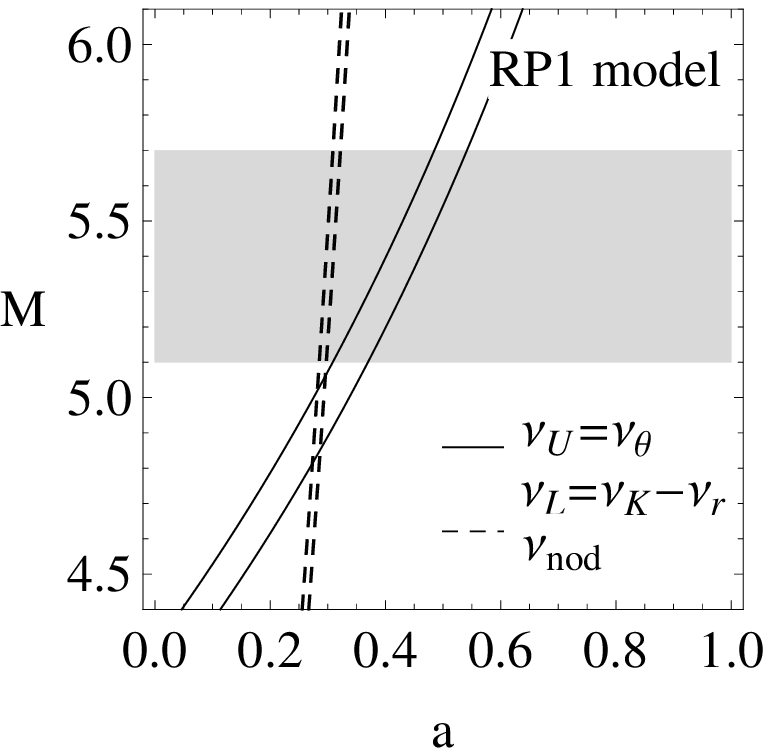}
\includegraphics[width=0.49\hsize]{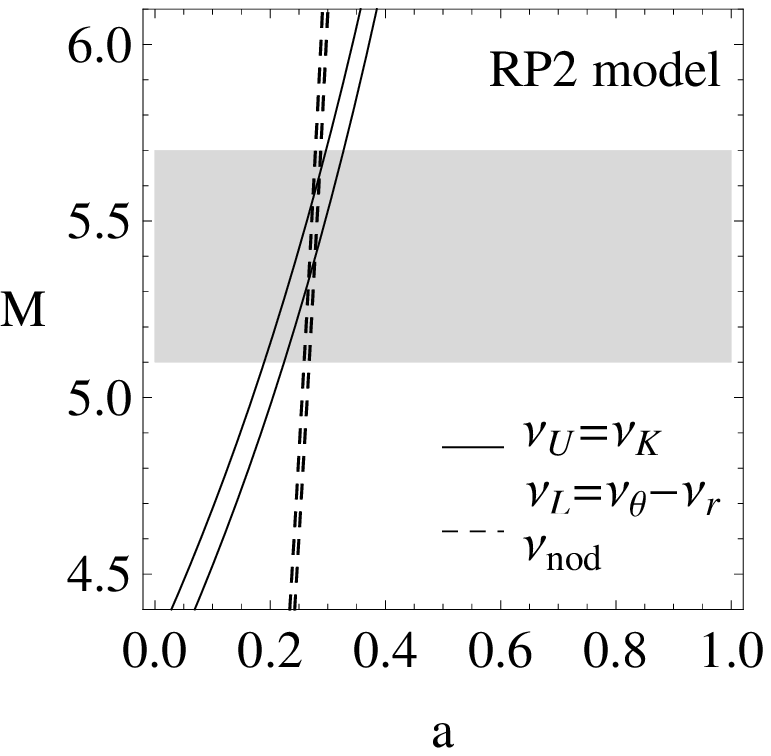}
\caption{
Restrictions on the parameters $M$ and $a$ given by the {\it RP, RP1}, and {\it RP2} models due to the QPO data simultaneously observed in the microquasar GRO~J1655-40. The solid lines are given by the $3\nu_{\rm L} \sim 2\nu_{\rm U}$ twin HF~QPOs resonance, while the dashed lines are obtained for the nodal frequency $\nu_{\rm nod}$ that explains the LF~QPOs at the same $r_{3:2}$ radius. The crossing of the twin HF QPO and the LF QPO limits implies the mass and spin of the black hole. The optical mass limit is shaded. 
\label{figRP}
}
\end{figure}

\begin{figure}
\includegraphics[width=0.49\hsize]{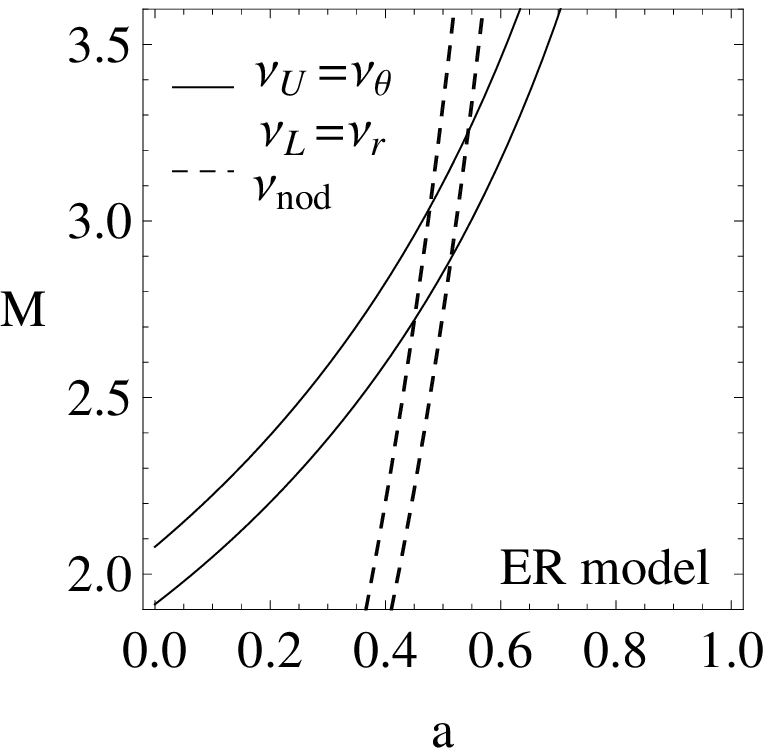}
\includegraphics[width=0.49\hsize]{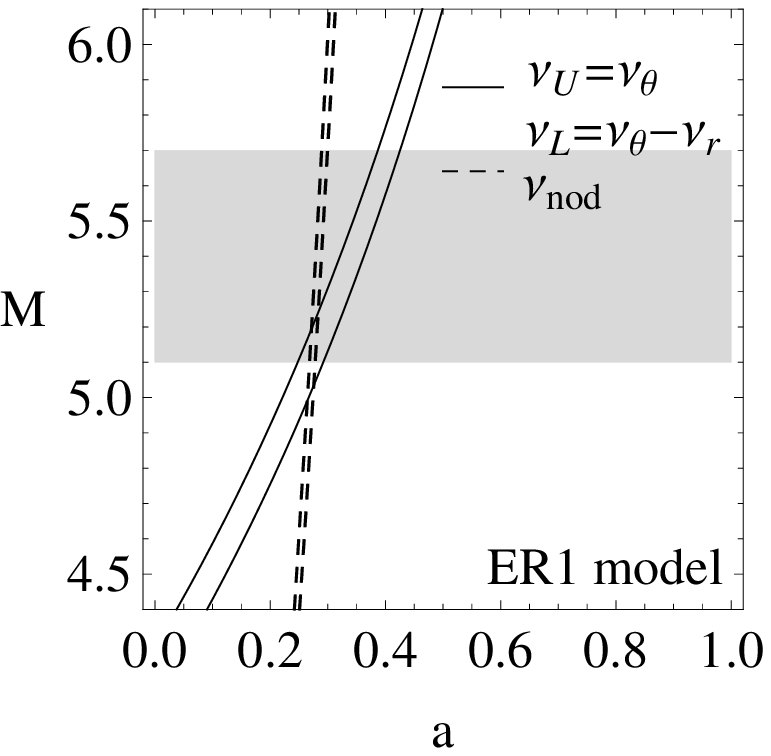}
\includegraphics[width=0.49\hsize]{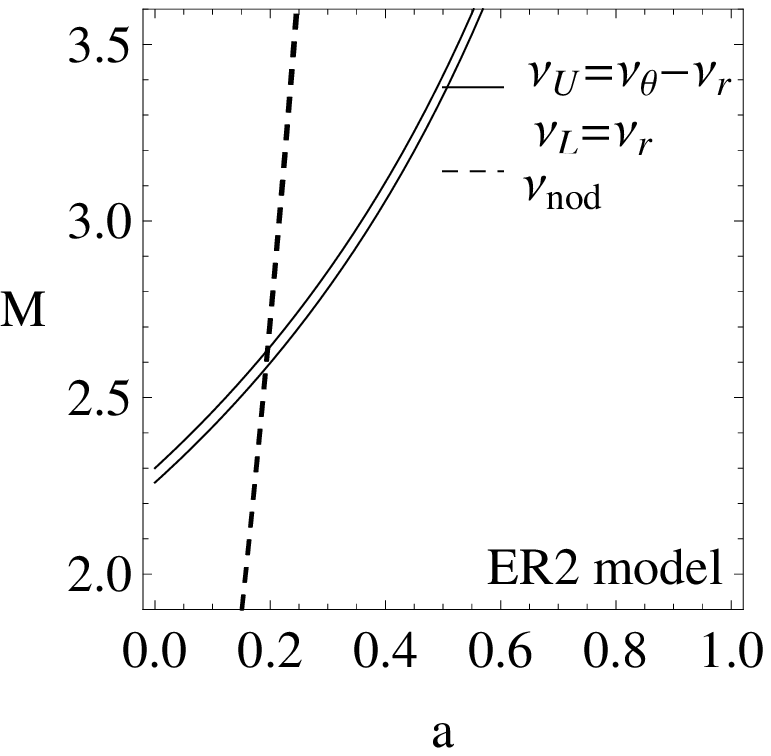}
\includegraphics[width=0.49\hsize]{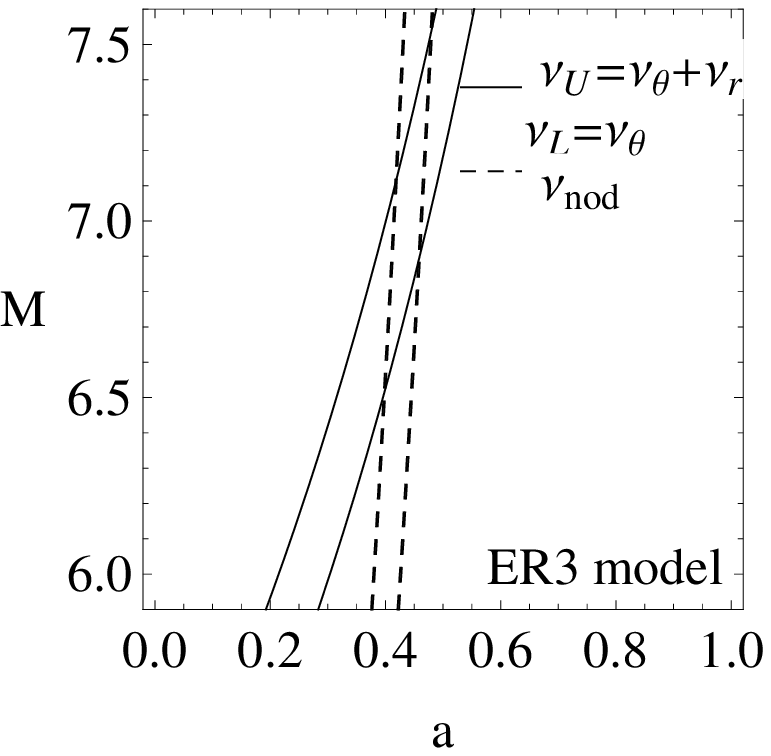}
\includegraphics[width=0.49\hsize]{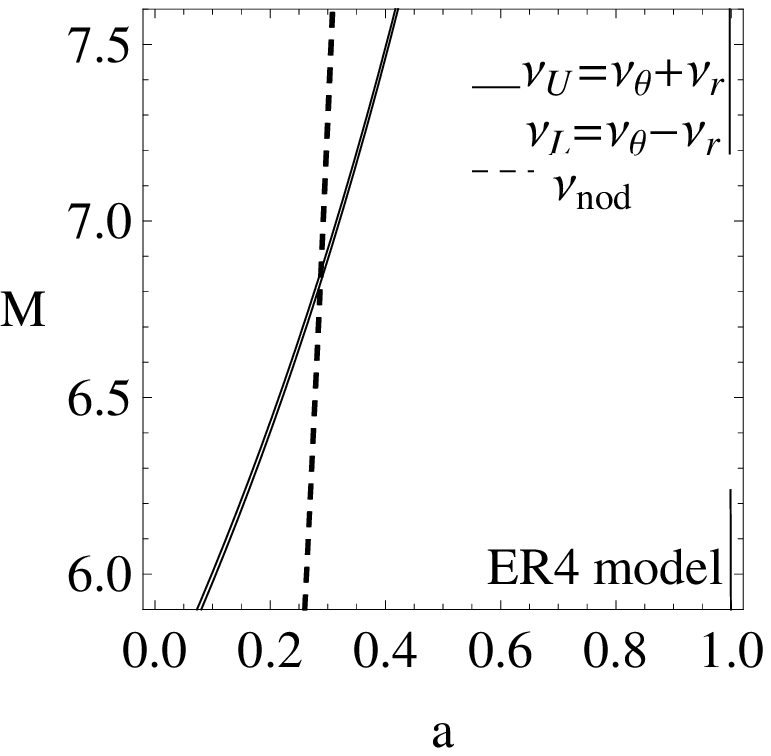}
\includegraphics[width=0.49\hsize]{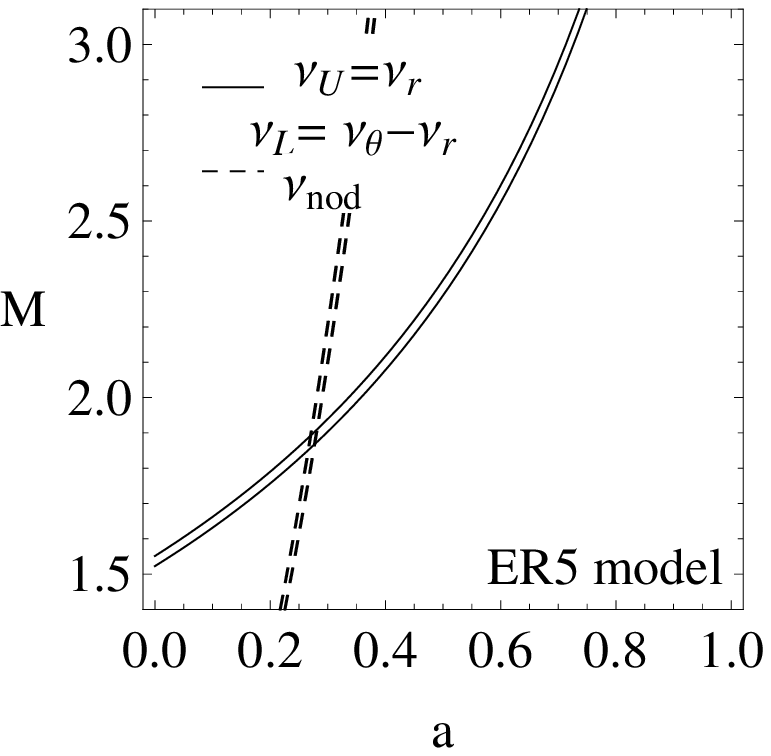}
\caption{
Restrictions on the parameters $M$ and $a$ given by the {\it ER, ER1 - ER5} models due to the QPO data simultaneously observed in the microquasar GRO~J1655-40. The solid lines are given by the $3\nu_{\rm L} \sim 2\nu_{\rm U}$ twin HF~QPOs resonance, while the dashed lines are obtained for the nodal frequency $\nu_{\rm nod}$ that explain the LF~QPOs at the same $r_{3:2}$ radius. The crossing of the twin HF QPO and the LF QPO limits implies the mass and spin of the black hole. The optical mass limit is shaded. 
\label{figRE}
}
\end{figure}

\begin{figure}
\includegraphics[width=0.49\hsize]{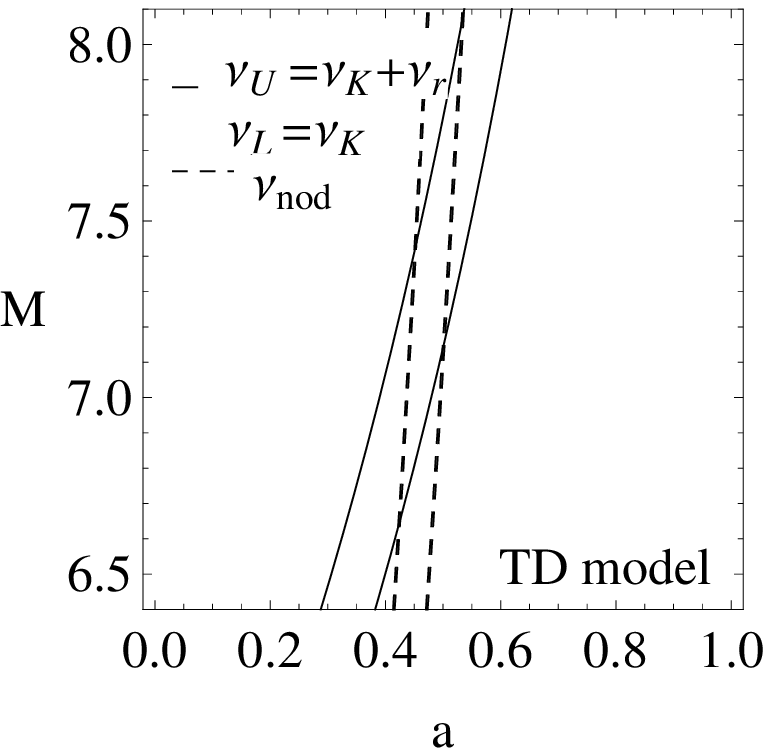}
\includegraphics[width=0.49\hsize]{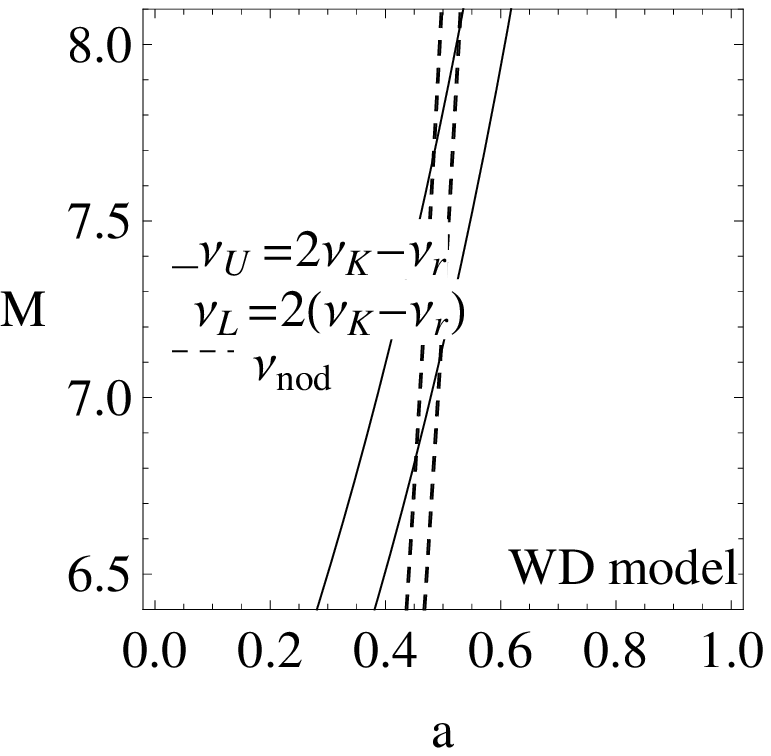}
\caption{
Restrictions on the parameters $M$ and $a$ given by  given by the {\it TD} and {\it TW} models due to the QPO data simultaneously observed in the microquasar GRO~J1655-40. 
The solid lines are given by the $3\nu_{\rm L} \sim 2\nu_{\rm U}$ twin HF~QPOs resonance, while the dashed lines are obtained for the nodal frequency $\nu_{\rm nod}$ that explain the LF~QPOs at the same $r_{3:2}$ radius. The crossing of the twin HF QPO and the LF QPO limits implies the mass and spin of the black hole. The optical mass limit is shaded. 
\label{figTDTW}
}
\end{figure}

\section{Matching the observed QPO frequencies}

Using the RP model of the twin HF QPOs including the nodal model of the LF QPO and the Monte Carlo technique of matching the models to the observations, the mass and spin of the GRO~J1655-40 black hole were established with high precision, $M=(5.31 \pm 0.07)~M_{\odot}$ and $a=0.290\pm0.003$ \citep{Mot-etal:2014a:MNRAS:}, in agreement with the mass limit given by independent optical measurements \citep{Bee-Pod:2002:MNRAS:}. 

We assume that the QPOs are governed by the geometry of the Kerr black hole when a unique relation $x^{\nu_{\rm U}(\mip,r,\theta)/\nu_{\rm L}(\mip,r,\theta)}(a;p)$ exists for each oscillation model based on the geodesic motion \citep{Stu-etal:2015:ACTA:}. Using the same frequency set as in \cite{Mot-etal:2014a:MNRAS:}, but a different frequency ratio technique of matching the models to the data, we test whether predictions of the standard twin HF QPO models extended by the nodal model of LF QPOs can be in agreement with the optical limit on the mass of the GRO~J1655-40 black hole. We assume that for the twin HF QPOs any frequency from the interval of allowed values of the upper (centroid) frequency can be combined with any frequency from the allowed interval of the lower frequency. The frequency ratio technique consists in the following succeeding steps that are the same for each of the selected oscillation models based on the frequencies governed by the geodesic quasi-circular motion in the Kerr geometry. Because of the same mass scaling of the orbital and epicyclic frequencies valid in the Kerr geometry, the frequency ratio method enables simple and effective matching to observational twin HF QPO data, as any combination of the orbital and epicyclic frequencies has identical mass scaling \citep{Stu-Kot-Tor:2013:ASTRA:}. \footnote{The frequency ratio method works for any frequency ratio of twin HF QPOs in any model of related twin oscillation modes with frequencies having the same mass scaling corresponding to the geodesic motion. The method can thus also work  for the external Hartle-Thorne geometry describing rotating neutron stars where all  three frequencies of the quasi-circular geodesic motion also have  the same mass scaling \cite{Stu-etal:2015:ACTA:}.} 

(i) We determine the range of frequency ratio $p$ of the measured upper and lower centroid frequencies of the twin HF QPOs with the related errors, e.g. $p_1 < p < p_2$. To find the radius-spin and mass-spin relations with errors implied by the errors of measured upper and lower twin HF QPO frequencies, it is enough to consider the frequency ratios at the edges of the allowed frequency ratio interval, i.e. at the minimal ratio $p=p_1$ and the maximal ratio $p=p_2$. These errors represent maximum errors (rather than  statistical errors).

(ii) We use the frequency ratio relation $a^{\nu_{\rm U}(\mip,r,\theta)/\nu_{\rm L}(\mip,r,\theta)}(x,p)$ for the maximum and minimum values of the frequency ratio $p$ and give the related dimensionless radius where the twin oscillations occur $x^{\nu_{\rm U}/\nu_{\rm L}}(a,p)$. \footnote{The solution of the equation $a=a^{\nu_{\rm U}/\nu_{\rm L}}(x,p)$ is unique in the Kerr black hole spacetimes \citep{Stu-Kot-Tor:2013:ASTRA:}.} For a given ratio $p$, the radius $x^{\nu_{\rm U}/\nu_{\rm L}}(a,p)$ is considered as a function of spin $a$ in the whole interval $0<a<1$.

(iii) The mass-spin relation due to the twin HF QPOs is adjusted by matching the upper model frequency to the upper value of the observed frequencies. For the chosen upper frequency and the radius $x^{\nu_{\rm U}/\nu_{\rm L}}(a,p)$ related to the spin $a$ by the preceding procedures, we determine the relation $M_{\rm HF}^{\nu_{\rm U}(\mip,r,\theta)/\nu_{\rm L}(\mip,r,\theta)}(a,p)$. We give the relations for $p=p_1$ and $p=p_2$. 

(iv) The restrictions from the nodal frequency model related to the LF QPO are given in the same way as for the twin HF QPOs. At each dimensionless radius predicted by the twin HF QPOs model, $x^{\nu_{\rm U}/\nu_{\rm L}}(a,p)$, we assume the occurrence of the observed low frequency QPO (at the edges of the interval given by the measurement error). We thus find the mass-spin relation related to the LF QPO using the radius-spin relation related to the twin HF QPOs because the nodal frequency can be simply expressed in the form 
\beq
        \nu_{nod}(M,a;p) = \frac{1}{2\pi M x_{HF}^{3/2}}(1+\frac{a}{x_{HF}^{3/2}})^{-1} 
                            [1 - (1 - \frac{4a}{x_{HF}^{3/2}} + \frac{3a^2}{x_{HF}^2})^{1/2}]
,\eeq 
where $x_{HF} = x^{\nu_{\rm U}/\nu_{\rm L}}(a,p)$ and is considered for the extremal values of ratio parameter $p$. The condition $\nu_{LFQPO} = \nu_{nod}$ then enables the determination of the mass spin relation corresponding to the LF QPO, $M_{\rm LF}^{\nu_{\rm U}(\mip,r,\theta)/\nu_{\rm L}(\mip,r,\theta)}(a,p)$. 

(v) Combining the restrictions implied by the twin HF QPOs and the LF QPO under assumption of their occurrence in a common radius, $M_{\rm HF}^{\nu_{\rm U}(\mip,r,\theta)/\nu_{\rm L}(\mip,r,\theta)}(a,p)$ and $M_{\rm LF}^{\nu_{\rm U}(\mip,r,\theta)/\nu_{\rm L}(\mip,r,\theta)}(a,p)$, we obtain restrictions on the black hole mass and spin for each of the considered twin HF QPO models. The errors in determining mass $M$ and spin $a$ of the black hole are given by the intersections of the mass-spin relations related to the twin HF QPOs and the LF QPO, which are governed by the statistical errors in the measured QPO centroid frequencies. The errors of $M$ and $a$ obtained this way are maximum  and slightly larger than those obtained in \cite{Mot-etal:2014a:MNRAS:}. 

Results of the numerical calculations for the RP, RP1, and RP2 models are presented in Figure \ref{figRP} for the ER and ER1-ER5 models in Figure \ref{figRE} and for the TD and WD models in Figure \ref{figTDTW}. The QPO limits on the black hole mass and spin are compared with the mass limit implied by the optical measurements. The ranges of the allowed values of the black hole mass and spin determined for all the selected twin HF QPO models combined with the nodal model of the LF QPOs are presented in Tab. \ref{tab1}. The range of allowed radii of simultaneous occurrence of both twin HF QPOs and the LF QPO is added. We also add  information related solely to the twin HF QPOs models;  we give the interval of mass parameter $M_{min}-M_{max}$ related to the whole interval of the black hole spin $0 < a < 1$, and the spin interval $a_{min} - a_{max}$ related to the mass restrictions given by the optical measurements that simply determine  the restrictions of the twin HF QPO models considered without relation to the LF QPO. 

Only three of the considered oscillation models match the optical mass limit. The RP model predicts $M =(5.3 \pm0.1)~M_{\odot}, a=0.286\pm0.004$, in agreement with the limits on mass and spin predicted by the Monte Carlo method \citep{Mot-etal:2014a:MNRAS:}. The so-called total precession RP2 model predicts $M=(5.5\pm0.1)~M_{\odot}, a=0.276\pm0.003$, while the resonance epicyclic model ER1 with the beat frequency $\nu_{-} = \nu_\mit - \nu_\mir = \nu_{\rm L}$ predicts $M=(5.1\pm0.1)~M_{\odot}, a=0.274\pm0.003$. The other oscillation models can be excluded. 

As already mentioned in \cite{Mot-etal:2014a:MNRAS:}, the predicted spin $a<0.3$ is much smaller than the estimates implied by the X-ray spectral analysis. The spectral continuum measurements predict $0.65<a<0.75$ \citep{Sha-etal:2006:ApJ:}, while the Fe-line profile measurements predict $0.94<a<0.98$ \citep{Mil-etal:2009:ApJ:};  we note  the clear discrepancy in black hole spin restrictions given by the two X-ray spectral methods. 
Since the spin restrictions implied by both the methods of the X-ray spectral analysis  contradict each other, they cannot be simultaneously matched by the geodesic models of the twin HF QPOs and the LF QPO.

We can also see in Tab. \ref{tab1}. that there is no model of twin HF QPOs that could match the spin limits of $0.65 < a < 0.75$ given by the X-ray spectral continuum measurements, while the spin limits $0.94 < a < 0.98$ given by the spectral profiled line measurements can be matched by the ER model, and by its ER5 variant; the ER2 variant only touches the spin interval from below. All the other models predict that the spin is too small to be matched to the X-ray spectral measurements. 

\begin{table*}
\begin{center}
\begin{tabular}{| l l l l l l l | l l |}
\hline
model & $\nu_U$ & $\nu_L$  & $\nu_{\rm low}$  & $M/M_{\odot}$ & $a$ & $r/r_{\rm g}$ & $(M_{\rm min}$--$M_{\rm max})/M_{\odot}$ & $a_{\rm min}$--$a_{\rm max}$ \\
\hline \hline
RP & $\nu_{\rm \mip}$                   & $\nu_{\rm \mip}-\nu_r $               &$\nu_\mip-\nu_\theta$ & $5.3\pm0.1$ & $0.286\pm0.004$ & $5.68\pm0.05$ &  4.1--17.8 & 0.22--0.38 \\
RP1 & $\nu_{\rm \theta}$ & $\nu_{\rm \mip} - \nu_r $  &$\nu_\mip-\nu_\theta$ & $4.9\pm0.1$ & $0.284\pm0.004$ & $5.81\pm0.05$ & 4.1--8.6 &  0.31--0.54 \\
RP2 & $\nu_{\rm \mip}$                  & $\nu_\theta -\nu_r $          &$\nu_\mip-\nu_\theta$ & $5.5\pm0.1$ & $0.276\pm0.003$ & $5.56\pm0.05$ & 4.1--13.4 &  0.19--0.33  \\
\hline 
ER  & $\nu_\theta$                                                      & $\nu_r $                                                &$\nu_\mip-\nu_\theta$ & $3.0\pm0.1$ & $0.496\pm0.020$ & $8.15\pm0.17$ & 1.9--7.1 & 0.87--0.96 \\
ER1 & $\nu_{\rm \theta}$                                & $\nu_{\rm \theta}-\nu_r $  &$\nu_\mip-\nu_\theta$ & $5.1\pm0.1$ & $0.274\pm0.003$ & $5.67\pm0.05$ & 4.1--11.2 & 0.25--0.43 \\
ER2 & $\nu_{\rm \theta}-\nu_{r}$ & $\nu_r $                                                                      &$\nu_\mip-\nu_\theta$  & $2.6\pm0.1$ & $0.194\pm0.002$ & $6.40\pm0.03$ & 2.3--6.6 & 0.87--0.07 \\
ER3 & $\nu_{\rm \theta}+\nu_{r}$ & $\nu_{\rm \theta} $                           &$\nu_\mip-\nu_\theta$ & $7.0\pm0.1$ & $0.439\pm0.022$ & $5.86\pm0.12$  & 4.8--16.3 & 0.00--0.24 \\
ER4 & $\nu_{\rm \theta}+\nu_{r}$ & $\nu_{\rm \theta}-\nu_{r}$ &$\nu_\mip-\nu_\theta$&  $6.8\pm0.1$ & $0.288\pm0.003$ & $5.22\pm0.02$ & 5.6--13.4 & 0.00--0.03 \\
ER5 & $\nu_r $                                                                  & $\nu_{\rm \theta}-\nu_{r}$ &$\nu_\mip-\nu_\theta$& $1.9\pm0.1$ & $0.274\pm0.005$ & $7.96\pm0.06$ & 1.5--5.5 & 0.98--1.00 \\
\hline
TD & $\nu_{\rm \mip} + \nu_r$                   & $\nu_{\rm \mip}$              &$\nu_\mip-\nu_\theta$                          &  $7.3\pm0.1$ & $0.477\pm0.026$ & $5.92\pm0.13$ & 4.8--17.3 & 0.00--0.24 \\
WD & $2\nu_{\rm \mip} - \nu_r$          & 2($\nu_{\rm \mip}-\nu_r) $ &$\nu_\mip-\nu_\theta$  & $7.4\pm0.3$ & $0.490\pm0.008$ & $5.94\pm0.11$ & 4.8--16.9 & 0.00--0.24 \\
\hline
\end{tabular}
\caption{
Restrictions on the parameters $M$ and $a$ of the black hole in the microquasar GRO~J1655-40 given by the geodesic QPO models applied to the simultaneously observed twin HF QPOs and LF QPO. The last two columns are related to the twin HF QPO models considered without the relation to the low frequency QPO. Different models are used, namely: relativistic precession and its variants (RP,RP1), total precession (RP2), parametric resonance and its variants (ER), tidal disruption (TD) and warped disc (WD). Dimensionless units are used with $r_{\rm g} = {\rm GM / c^2}$.
\label{tab1}
} 
\end{center}
\end{table*}

\section{Conclusions}

We have demonstrated that along with the RP model,  the so called total precession RP2 model and the forced resonance (beat frequency) epicyclic model ER1 can also  explain the twin HF QPOs simultaneously observed with the LF QPO in the microquasar GRO J1655-40 and predict the black hole mass in the range $5.1<M/M_{\odot}<5.5$, in agreement with the mass limit determined by the optical measurements. While the RP model predicts spin $a \sim 0.286$, the RP1 and EP1 models predict $a \sim 0.275$. 

All the models of the twin HF QPOs combined with the nodal model of the LF QPO, based on combinations of frequencies of the geodesic quasi-circular motion, predict spin $a < 0.3$, clearly contradicting the spin estimates due to the spectral measurements giving $a>0.65$, as mentioned in \cite{Mot-etal:2014a:MNRAS:} for the RP model. \footnote{However, it should be noted that the RP model applied to the QPOs in the source XTE~J1550-564, and combined with the optical mass limit, implies a black hole spin consistent with the X-ray spectral measurements \citep{Mot-etal:2014b:MNRAS:}.} 
If the black hole spin has to be in agreement with the spin spectral continuum measurements restrictions $0.65<a<0.75$ \citep{Sha-etal:2006:ApJ:}, no geodesic model of twin HF QPOs can match the observed $3:2$ twin high-frequency modes alone.

It remains to be determined which of the QPO models is the correct one, if any. All the models are based on the frequencies of the quasi-circular geodesic motion when gravity is considered to be the only relevant factor, but inclusion of non-geodesic effects, related to plasma phenomena or fluid pressure for example, could introduce relevant modifications to the QPO models that have to be addressed in future studies. 

\section*{Acknowledgments}

The authors would like to thank the institutional support of the Faculty of Philosophy and Science of the Silesian University in Opava, and the Albert Einstein Centre for Gravitation and Astrophysics supported by the~Czech Science Foundation Grant No. 14-37086G. 


\def\prc{Phys. Rev. C}
\def\pre{Phys. Rev. E}
\def\prd{Phys. Rev. D}
\def\jcap{Journal of Cosmology and Astroparticle Physics}
\def\apss{Astrophysics and Space Science}
\def\mnras{Mon. Not. R. Astron Soc.}
\def\apj{The Astrophysical Journal}
\def\aap{Astron. Astrophys.}
\def\actaa{Acta Astronomica}
\def\pasj{Publications of the Astronomical Society of Japan}
\def\apjl{Astrophysical Journal Letters}
\def\pasa{Publications Astronomical Society of Australia}
\def\nat{Nature}
\def\physrep{Phys. Rep.}
\def\araa{Annu. Rev. Astron. Astrophys.}
\def\apjs{The Astrophysical Journal Supplement}

\bibliographystyle{aa}

\begin{thebibliography}{56}
\expandafter\ifx\csname natexlab\endcsname\relax\def\natexlab#1{#1}\fi

\bibitem[{{Abramowicz} {et~al.}(1978){Abramowicz}, {Jaroszynski}, \&
  {Sikora}}]{Abr-etal:1978:ASTRA:}
{Abramowicz}, M., {Jaroszynski}, M., \& {Sikora}, M. 1978, \aap, 63, 221

\bibitem[{{Abramowicz} \& {Klu{\'z}niak}(2001)}]{Abr-Klu:2001:ASTRA:}
{Abramowicz}, M.~A. \& {Klu{\'z}niak}, W. 2001, \aap, 374, L19

\bibitem[{{Aliev} \& {Galtsov}(1981)}]{Ali-Gal:1981:GRG:}
{Aliev}, A.~N. \& {Galtsov}, D.~V. 1981, General Relativity and Gravitation,
  13, 899

\bibitem[{{Bardeen} {et~al.}(1972){Bardeen}, {Press}, \&
  {Teukolsky}}]{Bar-Pre-Teu:1972:ApJ:}
{Bardeen}, J.~M., {Press}, W.~H., \& {Teukolsky}, S.~A. 1972, \apj, 178, 347

\bibitem[{{Beer} \& {Podsiadlowski}(2002)}]{Bee-Pod:2002:MNRAS:}
{Beer}, M.~E. \& {Podsiadlowski}, P. 2002, \mnras, 331, 351

\bibitem[{{Belloni} {et~al.}(2007){Belloni}, {M{\'e}ndez}, \&
  {Homan}}]{Bel-Men-Hom:2007:MONNRS:}
{Belloni}, T., {M{\'e}ndez}, M., \& {Homan}, J. 2007, \mnras, 376, 1133

\bibitem[{{Boshkayev} {et~al.}(2014){Boshkayev}, {Bini}, {Rueda}, {Geralico},
  {Muccino}, \& {Siutsou}}]{Bosh-etal:2014:GraCos:}
{Boshkayev}, K., {Bini}, D., {Rueda}, J., {et~al.} 2014, Gravitation and
  Cosmology, 20, 233

\bibitem[{{Bursa}(2005)}]{Bur:2005:RAG:}
{Bursa}, M. 2005, in RAGtime 6/7: Workshops on black holes and neutron stars,
  ed. S.~{Hled{\'{\i}}k} \& Z.~{Stuchl{\'{\i}}k}, 39--45

\bibitem[{{Casella} {et~al.}(2005){Casella}, {Belloni}, \&
  {Stella}}]{Cas-Bel-Ste:2005:ApJ:}
{Casella}, P., {Belloni}, T., \& {Stella}, L. 2005, \apj, 629, 403

\bibitem[{{{\v C}ade{\v z}} {et~al.}(2008){{\v C}ade{\v z}}, {Calvani}, \&
  {Kosti{\'c}}}]{Cad-Cal-Kos:2008:ASTRA:}
{{\v C}ade{\v z}}, A., {Calvani}, M., \& {Kosti{\'c}}, U. 2008, \aap, 487, 527

\bibitem[{{Feroci} {et~al.}(2012){Feroci}, {den Herder}, {Bozzo}, {Barret},
  {Brandt}, {Hernanz}, {van der Klis}, {Pohl}, {Santangelo}, {Stella}, \&
  et~al.}]{Fer-etal:2012:ExpAstr:}
{Feroci}, M., {den Herder}, J.~W., {Bozzo}, E., {et~al.} 2012, Experimental
  Astronomy, 34, 415

\bibitem[{{Kato}(2004)}]{Kat:2004:PASJ:}
{Kato}, S. 2004, \pasj, 56, 905

\bibitem[{{Kato}(2008)}]{Kat:2008:PASJ:}
{Kato}, S. 2008, \pasj, 60, 889

\bibitem[{{Kato} \& {Fukue}(1980)}]{Kat-Fuk:1980:PASJ:}
{Kato}, S. \& {Fukue}, J. 1980, \pasj, 32, 377

\bibitem[{{Kato} {et~al.}(1998){Kato}, {Fukue}, \&
  {Mineshige}}]{Kat-Fuk-Min:1998:BHAccDis:}
{Kato}, S., {Fukue}, J., \& {Mineshige}, S., eds. 1998, {Black-hole accretion
  disks}

\bibitem[{{Kosti{\'c}} {et~al.}(2009){Kosti{\'c}}, {{\v C}ade{\v z}},
  {Calvani}, \& {Gomboc}}]{Kos-etal:2009:ASTRA:}
{Kosti{\'c}}, U., {{\v C}ade{\v z}}, A., {Calvani}, M., \& {Gomboc}, A. 2009,
  \aap, 496, 307

\bibitem[{{Kotrlov{\'a}} {et~al.}(2014){Kotrlov{\'a}}, {T{\"o}r{\"o}k}, {{\v
  S}r{\'a}mkov{\'a}}, \& {Stuchl{\'{\i}}k}}]{Kot-etal:2014:ASTRA:}
{Kotrlov{\'a}}, A., {T{\"o}r{\"o}k}, G., {{\v S}r{\'a}mkov{\'a}}, E., \&
  {Stuchl{\'{\i}}k}, Z. 2014, \aap, 572, A79

\bibitem[{{Kozlowski} {et~al.}(1978){Kozlowski}, {Jaroszynski}, \&
  {Abramowicz}}]{Koz-etal:1977:ASTRA:}
{Kozlowski}, M., {Jaroszynski}, M., \& {Abramowicz}, M.~A. 1978, \aap, 63, 209

\bibitem[{{Landau} \& {Lifshitz}(1969)}]{Lan-Lif:1969:Mech:}
{Landau}, L.~D. \& {Lifshitz}, E.~M. 1969, {Mechanics} (Oxford: Pergamon Press)

\bibitem[{{Lin} {et~al.}(2011){Lin}, {Boutelier}, {Barret}, \&
  {Zhang}}]{Lin-etal:2011:ApJ:}
{Lin}, Y.-F., {Boutelier}, M., {Barret}, D., \& {Zhang}, S.-N. 2011, \apj, 726,
  74

\bibitem[{{Miller} {et~al.}(2009){Miller}, {Cackett}, \&
  {Reis}}]{Mil-etal:2009:ApJ:}
{Miller}, J.~M., {Cackett}, E.~M., \& {Reis}, R.~C. 2009, \apjl, 707, L77

\bibitem[{{Miller} {et~al.}(1998){Miller}, {Lamb}, \&
  {Psaltis}}]{Mil-Lam-Psa:1998:ApJ:}
{Miller}, M.~C., {Lamb}, F.~K., \& {Psaltis}, D. 1998, \apj, 508, 791

\bibitem[{{Montero} \& {Zanotti}(2012)}]{Mon-Zan:2012:MNRAS:}
{Montero}, P.~J. \& {Zanotti}, O. 2012, \mnras, 419, 1507

\bibitem[{{Motta} {et~al.}(2012){Motta}, {Homan}, {Mu{\~n}oz Darias},
  {Casella}, {Belloni}, {Hiemstra}, \& {M{\'e}ndez}}]{Mot-etal:2012:MNRAS:}
{Motta}, S., {Homan}, J., {Mu{\~n}oz Darias}, T., {et~al.} 2012, \mnras, 427,
  595

\bibitem[{{Motta} {et~al.}(2014{\natexlab{a}}){Motta}, {Belloni}, {Stella},
  {Mu{\~n}oz-Darias}, \& {Fender}}]{Mot-etal:2014a:MNRAS:}
{Motta}, S.~E., {Belloni}, T.~M., {Stella}, L., {Mu{\~n}oz-Darias}, T., \&
  {Fender}, R. 2014{\natexlab{a}}, \mnras, 437, 2554

\bibitem[{{Motta} {et~al.}(2014{\natexlab{b}}){Motta}, {Mu{\~n}oz-Darias},
  {Sanna}, {Fender}, {Belloni}, \& {Stella}}]{Mot-etal:2014b:MNRAS:}
{Motta}, S.~E., {Mu{\~n}oz-Darias}, T., {Sanna}, A., {et~al.}
  2014{\natexlab{b}}, \mnras, 439, L65

\bibitem[{{Mukhopadhyay}(2009)}]{Muk:2009:ApJ:}
{Mukhopadhyay}, B. 2009, \apj, 694, 387

\bibitem[{{Nayfeh} \& {Mook}(1979)}]{Nay-Moo:1979:NonOscilations:}
{Nayfeh}, A.~H. \& {Mook}, D.~T. 1979, {Nonlinear oscillations} (New York :
  Wiley)

\bibitem[{{Novikov} \& {Thorne}(1973)}]{Nov-Tho:1973:BlaHol:}
{Novikov}, I.~D. \& {Thorne}, K.~S. 1973, in Black Holes (Les Astres Occlus),
  ed. C.~{Dewitt} \& B.~S. {Dewitt}, 343--450

\bibitem[{{Page} \& {Thorne}(1974)}]{Pag-Tho:1974:ApJ:}
{Page}, D.~N. \& {Thorne}, K.~S. 1974, \apj, 191, 499

\bibitem[{{Pappas}(2012)}]{Papp:2012:MONNR:}
{Pappas}, G. 2012, \mnras, 422, 2581

\bibitem[{{Rezzolla} {et~al.}(2003){Rezzolla}, {Yoshida}, {Maccarone}, \&
  {Zanotti}}]{Rez-etal:2003:MNRAS:}
{Rezzolla}, L., {Yoshida}, S., {Maccarone}, T.~J., \& {Zanotti}, O. 2003,
  \mnras, 344, L37

\bibitem[{{Shafee} {et~al.}(2006){Shafee}, {McClintock}, {Narayan}, {Davis},
  {Li}, \& {Remillard}}]{Sha-etal:2006:ApJ:}
{Shafee}, R., {McClintock}, J.~E., {Narayan}, R., {et~al.} 2006, \apjl, 636,
  L113

\bibitem[{{Stefanov}(2014)}]{Ste:2014:MNRAS:}
{Stefanov}, I.~Z. 2014, \mnras, 444, 2178

\bibitem[{{Stella} \& {Vietri}(1998)}]{Ste-Vie:1998:ApJ:}
{Stella}, L. \& {Vietri}, M. 1998, \apjl, 492, L59

\bibitem[{{Stella} \& {Vietri}(1999)}]{Ste-Vie:1999:PHYSRL:}
{Stella}, L. \& {Vietri}, M. 1999, Physical Review Letters, 82, 17

\bibitem[{{Stella} {et~al.}(1999){Stella}, {Vietri}, \&
  {Morsink}}]{Ste-Vie-Mor:1999:ApJ:}
{Stella}, L., {Vietri}, M., \& {Morsink}, S.~M. 1999, \apjl, 524, L63

\bibitem[{{Strohmayer}(2001)}]{Str:2001:ApJ:}
{Strohmayer}, T.~E. 2001, \apjl, 552, L49

\bibitem[{{Stuchl{\'{\i}}k}(1980)}]{Stu:1980:BAC:}
{Stuchl{\'{\i}}k}, Z. 1980, Bulletin of the Astronomical Institutes of
  Czechoslovakia, 31, 129

\bibitem[{{Stuchl{\'{\i}}k} {et~al.}(2011){Stuchl{\'{\i}}k}, {Kotrlov{\'a}}, \&
  {T{\"o}r{\"o}k}}]{Stu-Kot-Tor:2011:ASTRA:}
{Stuchl{\'{\i}}k}, Z., {Kotrlov{\'a}}, A., \& {T{\"o}r{\"o}k}, G. 2011, \aap,
  525, A82

\bibitem[{{Stuchl{\'{\i}}k} {et~al.}(2012){Stuchl{\'{\i}}k}, {Kotrlov{\'a}}, \&
  {T{\"o}r{\"o}k}}]{Stu-Kot-Tor:2012:ACTA:}
{Stuchl{\'{\i}}k}, Z., {Kotrlov{\'a}}, A., \& {T{\"o}r{\"o}k}, G. 2012, \actaa,
  62, 389

\bibitem[{{Stuchl{\'{\i}}k} {et~al.}(2013){Stuchl{\'{\i}}k}, {Kotrlov{\'a}}, \&
  {T{\"o}r{\"o}k}}]{Stu-Kot-Tor:2013:ASTRA:}
{Stuchl{\'{\i}}k}, Z., {Kotrlov{\'a}}, A., \& {T{\"o}r{\"o}k}, G. 2013, \aap,
  552, A10

\bibitem[{{Stuchl{\'{\i}}k} \& {Schee}(2010)}]{Stu-Sche:2010:CLAQG:}
{Stuchl{\'{\i}}k}, Z. \& {Schee}, J. 2010, Classical and Quantum Gravity, 27,
  215017

\bibitem[{{Stuchl{\'{\i}}k} \& {Schee}(2012)}]{Stu-Sche:2012:CLAQG:}
{Stuchl{\'{\i}}k}, Z. \& {Schee}, J. 2012, Classical and Quantum Gravity, 29,
  065002

\bibitem[{{Stuchl{\'{\i}}k} {et~al.}(2009){Stuchl{\'{\i}}k}, {Slan{\'y}}, \&
  {Kov{\'a}{\v r}}}]{Stu-etal:2009:CLAQG:}
{Stuchl{\'{\i}}k}, Z., {Slan{\'y}}, P., \& {Kov{\'a}{\v r}}, J. 2009, Classical
  and Quantum Gravity, 26, 215013

\bibitem[{{Stuchlik} {et~al.}(2015){Stuchlik}, {Urbanec}, {Kotrlova}, {Torok},
  \& {Goluchova}}]{Stu-etal:2015:ACTA:}
{Stuchlik}, Z., {Urbanec}, M., {Kotrlova}, A., {Torok}, G., \& {Goluchova}, K.
  2015, Acta Astronomica, 65, 169

\bibitem[{{T{\"o}r{\"o}k} {et~al.}(2005){T{\"o}r{\"o}k}, {Abramowicz},
  {Klu{\'z}niak}, \& {Stuchl{\'{\i}}k}}]{Tor-etal:2005:ASTRA:}
{T{\"o}r{\"o}k}, G., {Abramowicz}, M.~A., {Klu{\'z}niak}, W., \&
  {Stuchl{\'{\i}}k}, Z. 2005, \aap, 436, 1

\bibitem[{{T{\"o}r{\"o}k} {et~al.}(2010){T{\"o}r{\"o}k}, {Bakala}, {{\v
  S}r{\'a}mkov{\'a}}, {Stuchl{\'{\i}}k}, \& {Urbanec}}]{Tor-etal:2010:ApJ:}
{T{\"o}r{\"o}k}, G., {Bakala}, P., {{\v S}r{\'a}mkov{\'a}}, E.,
  {Stuchl{\'{\i}}k}, Z., \& {Urbanec}, M. 2010, \apj, 714, 748

\bibitem[{{T{\"o}r{\"o}k} {et~al.}(2012){T{\"o}r{\"o}k}, {Bakala}, {{\v
  S}r{\'a}mkov{\'a}}, {Stuchl{\'{\i}}k}, {Urbanec}, \&
  {Goluchov{\'a}}}]{Tor-etal:2012:ApJ:}
{T{\"o}r{\"o}k}, G., {Bakala}, P., {{\v S}r{\'a}mkov{\'a}}, E., {et~al.} 2012,
  \apj, 760, 138

\bibitem[{{T{\"o}r{\"o}k} {et~al.}(2011){T{\"o}r{\"o}k}, {Kotrlov{\'a}}, {{\v
  S}r{\'a}mkov{\'a}}, \& {Stuchl{\'{\i}}k}}]{Tor-etal:2011:ASTRA:}
{T{\"o}r{\"o}k}, G., {Kotrlov{\'a}}, A., {{\v S}r{\'a}mkov{\'a}}, E., \&
  {Stuchl{\'{\i}}k}, Z. 2011, \aap, 531, A59

\bibitem[{{T{\"o}r{\"o}k} \& {Stuchl{\'{\i}}k}(2005)}]{Tor-Stu:2005:ASTRA:}
{T{\"o}r{\"o}k}, G. \& {Stuchl{\'{\i}}k}, Z. 2005, \aap, 437, 775

\bibitem[{{Zanotti} {et~al.}(2005){Zanotti}, {Font}, {Rezzolla}, \&
  {Montero}}]{Zan-etal:2005:MNRAS:}
{Zanotti}, O., {Font}, J.~A., {Rezzolla}, L., \& {Montero}, P.~J. 2005, \mnras,
  356, 1371

\bibitem[{{Zhang} {et~al.}(2006){Zhang}, {Yin}, {Zhao}, {Zhang}, \&
  {Song}}]{Zhang-etal:2006:MONNR:}
{Zhang}, C.~M., {Yin}, H.~X., {Zhao}, Y.~H., {Zhang}, F., \& {Song}, L.~M.
  2006, \mnras, 366, 1373

\end{thebibliography}

\end{document}